\documentclass[%
reprint,
superscriptaddress,
amsmath,amssymb,
prb,
]{revtex4-1}
\usepackage[colorlinks,linkcolor=blue, citecolor=blue, urlcolor=blue]{hyperref}%pour colorer liste figures, ref...ET CLIQUER DESSUS dans le pdf
\usepackage{graphicx}% Include figure files
\usepackage{dcolumn}% Align table columns on decimal point
\usepackage{bm}% bold math
\usepackage{subfigure}
\usepackage{xcolor}
\newcommand\refeq[1]{Eq. \eqref{#1}}

\begin{document}

\preprint{APS/123-QED}

\title{Composition dependence of radiation induced patterns in non miscible alloys }%

\author{D. Simeone}%
\affiliation{CEA/DEN/DMN/SRMA/LA2M-LRC CARMEN, CEA, Universit\'e Paris-Saclay, F-91191, Gif-sur-Yvette, France, CNRS/ECP/UMR 8085, Grande voie des vignes, Chatenay Malabry France}%

\author{V. Pontikis}%
\affiliation{CEA/DRF/IRAMIS/LSI, CEA, Universit\'e Paris-Saclay, F-91191, Gif-sur-Yvette, France}%

\author{L. Luneville}%
\affiliation{CEA/DEN/DM2S/SERMA/LLPR-LRC CARMEN, CEA, Universit\'e Paris-Saclay, F-91191, Gif-sur-Yvette, France,CNRS/ECP/UMR 8085, Grande voie des vignes, Chatenay Malabry France France}%

\date{\today}

\begin{abstract}
{We present a theoretical approach exhaustively predicting the variety of steady-state shapes emerging under irradiation in thermodynamically unstable binary mixtures. We show that stripes or honeycomb structures are controlled  not only by the two classical irradiation parameters: the irradiation flux and the temperature, but also by the nominal composition of the mixture. A rationale is thereby established for the results found in the literature. Moreover, the present developments lead to a simple methodology for predicting irradiation patterning without solving any evolution equation. It is foreseen, that this hands-on method will allow preparing materials with desired properties stemming from metastable irradiation microstructures produced on demand.}
\end{abstract}

\pacs{Valid PACS appear here}

\maketitle{}
%%%%%%%%%%%

In this work, we show that steady-state shapes emerging under irradiation in thermodynamically unstable binary mixtures can be exhaustively predicted by a mean-field analytic approach without explicitly solving the evolution equation and that the main parameters controlling the microstructure are, the irradiation flux, $\phi$, the temperature, $T$, and the nominal composition of the mixture $\overline{c}$. Beneficial effects of this twofold achievement are (i) the rational classification of results existing in the literature~\cite{Adda75, Barbu80, Enrique00, Simeone13, Demange15} within a generic phase diagram revealing the link between steady-states and the above listed control parameters and (ii) the setup of a practical method for predicting the steady-state microstructures forming under irradiation in any decomposing binary mixture provided the triplet of values ($\overline{c}$,$\phi$,$T$) is specified. Thereby credit is given to the perspective of preparing materials with the desired properties via irradiation-driven tailoring of their microstructures.

In the following,  a succinct description is first given of the theoretical framework underlying this work, followed by the presentation of the additional developments we have made leading to the main findings summarized above. For the illustration purpose, the developed practical method has been applied to a random solid solution, $Ag_{0.39}Cu_{0.61}$, evolving under a flux, $\phi=6 \times 10^{12} cm^{-2}s^{-1}$, of 1 MeV Kr ions at T=440 K and numerically modelled in two dimensions (2D) for the sake of simplicity. Finally, the results are discussed and briefly compared to these found in the literature.

The time evolution of a decomposing binary mixture under irradiation additively combines the re-distribution of species via thermal diffusion enhanced by radiation-induced mobile point defects with average mobility, $\Gamma_{th}(T,\phi)$~\cite{Subram93} and atom relocation triggered by ballistic collisions between incident ions and the matrix yielding the mobility, $\Gamma_{irr}(\phi)$~\cite{Enrique00,Simeone13}. At the mesoscale, the local composition $c(\mathbf{r},t)$ of one species A of the alloy,  is usually described by the conserved order parameter $\eta(\mathbf{r},t)=c(\mathbf{r},t)-c_0$, where, $c_0$, is the composition of species A at the critical temperature~\cite{Subram93}.\\ Focusing on the low temperature range, thus neglecting noise effects, the thermal evolution of the decomposing mixture is given by~\cite{Simeone13,Sigmund81}:
\begin{equation}
\label{eq_CH}
{\frac{\partial \eta(\mathbf{r},t)}{\partial t}}\Biggr\rvert_{th}=\Gamma_{th}(T,\phi) \nabla^2 \frac{\delta F[\eta]}{\delta \eta }
\end{equation}
where $F[\eta]=\int f(\eta(\mathbf{r},t) d\mathbf{r}$ represents the free energy and determines the chemical affinity of species in the mixture.  In this expression, the free energy density $f(\eta(\mathbf{r},t))$ is represented by a Landau fourth order expansion $\frac{a_2}{2}{\eta(\mathbf{r},t)}^2+\frac{a_3}{3}{\eta(\mathbf{r},t)}^3+\frac{a_4}{4}{\eta(\mathbf{r},t)}^4$ appropriately describing first-order transformations (negative value of $a_3$) such the spinodal decomposition of mixtures, of central interest in this work~\cite{Khatchaturyan83, Toledano96}. Spatial heterogeneity  of $\eta(\mathbf{r},t)$ is represented by adding to $F[\eta]$ the Ginzburg term, $\int \kappa |\nabla \eta(\mathbf{r},t)|^2 d\mathbf{r}$ where $\kappa>0$ relates to the energetic cost of interfaces forming between separating phases~\cite{Khatchaturyan83}.\\
On the other hand, displacements
 of atoms induced by ballistic mixing under irradiation are modelled via~\cite{GrasMarti81,Simeone10}:
\begin{equation}
\label{eq_irr}
{\frac{\partial \eta(\mathbf{r},t)}{\partial t}}\Biggr\rvert_{irr}=\Gamma_{irr}(\phi) \left[ \int p_{R}(\mathbf{r-r'})
\eta(\mathbf{r'},t)dr'-\eta(\mathbf{r'},t) \right]
\end{equation}
with, $p_{R}(\mathbf{r})$, the probability density of atom relocation in displacement cascades and $R$ the mean free path of relocated atoms~\cite{Enrique00,Simeone13}. 
%%%%%%%%%%%%%%%%%%%%%%%%%%%%%%%%%%%%%%%%%%%%%%%%%%%%%%%%%%
\begin{figure}[ht]
\centering
\includegraphics[width=8cm]{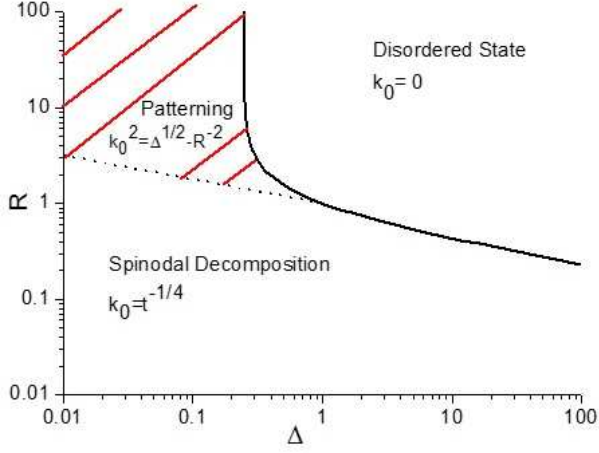}
 \caption{Phase diagram of steady states produced under irradiation in non miscible binary solutions irrespective to their nominal compositions. At low temperatures ($\Delta<1$) and large relocation distances ($R>1$), the patterned steady states emerging with characteristic wave vector $k_0=(\sqrt{\Delta}-R^{-2})^{1/2}$, do not appear in the classical thermodynamic phase diagram. Full line: limit between ordered and disordered states (\refeq{eq_dmax}), dashed line: limit between spinodal decomposition and patterning domain (\refeq{eq_dmin}) }
\label{fig_pattern} 
\end{figure}
%%%%%%%%%%%%%%%%%%%%%%%%%%%%%%%%%%%%%%%%%%%%%%%%%%%%%%%%%%
Considering the functional, $\mathcal{L}[\eta]=F[\eta] + \frac{\Delta(\phi,T)}{2}  G[\eta]$ where the kinetic enhancement factor, ${\Delta(\phi,T)=\frac{\Gamma_{irr}(\phi)}{\Gamma_{th}(\phi,T)}}$, and $G(\eta(\mathbf{r},t))=\int \int  \eta(\mathbf{r},t) g(\mathbf{r}-\mathbf{r'}) \eta(\mathbf{r'},t) d\mathbf{r} d\mathbf{r'}$ with $\nabla^2 g = p_R - \delta$~\cite{Enrique00, Luneville16},  the global evolution of the mixture is conveniently described by~\cite{Enrique00, Luneville16}:
\begin{equation}
\label{eq_Lyapounov}
{\frac{\partial \eta(\mathbf{r},t)}{\partial t}}\Biggr\rvert_{tot} 
= \Gamma_{th}(T,\phi) \nabla^2 \left[\frac{\partial \mathcal{L}(\eta(\mathbf{r},t))}{\partial \eta(\mathbf{r},t)}\right]
\end{equation}
\refeq{eq_Lyapounov} governs the evolution of the microstructure. Computing $\eta$ in large systems, the first and the third space derivatives of $\eta$ vanish at the system boundaries~\cite{Demange15}.
 The always positiveness and monotonous time decreasing of $\mathcal{L[\eta]}$~\cite{Demange15} insures that $\mathcal{L[\eta]}$ is a Lyapounov functional representing the effective free energy of the system. Thus, minima of  $\mathcal{L[\eta]}$ correspond to all steady states solutions of \refeq{eq_Lyapounov}. For a decomposing under irradiation binary solution, these can be classified  in a pseudo phase diagram spanned by, R and $\Delta$, the species relocation and the kinetic enhancement factors respectively~\cite{Enrique00, Luneville16}, as is shown in Fig.~\ref{fig_pattern}.
However, at this stage, in the patterning domain (hatched zone in Fig.~\ref{fig_pattern}), specific information about the symmetry elements and the distribution in space of emerging phases as well as their composition is not available, whereas it is intuitively foreseen that the  nominal composition of the solution, $\overline{c}$, should be the selection factor determining the kind of emerging patterns. 
\newline
\newline
The main objective of the present work is to establish a link between $\overline{c}$, the composition and the symmetry elements of the spatial distribution of irradiation patterns, proving thereby that Fig.~\ref{fig_pattern} represents a 2D projection of the complete 3D phase diagram including as third axis the nominal composition of the mixture. 

For estimating this last, it is first worth remarking that in the long-time limit, the structure factor $S(k,t)$~\cite{Khatchaturyan83} of the evolving patterns is sharply peaked at the wave vector with modulus $k_0$ minimizing $\mathcal{L}$, whereas its width evolves with time as, $t^{-\frac{1}{4}}$~\cite{Glotzer95, Simeone13}.
Therefore, the long-time limit of, $D(k)=S(k,\infty)^{-1}$, is reasonably approximated by the following second-order expansion~\cite{Jin06}:
\begin{eqnarray}
D(k) &\approx& S^{-1}(k_0)-\frac{S"(k_0)}{8k_0^2 S(k_0)}(k^2-k_0^2)^2 \nonumber \\
&\approx& D(k_0)+\frac{D"(k_0)}{8k_0^2}(k^2-k_0^2)^2
\label{eq_Dk}
\end{eqnarray}
with $D"(k_0)$ the second derivative of $D(k)$ at $k_0$. As $k_0$ is a maximum of $S$, $D"(k_0)$ is always positive.

\begin{figure}[ht]
\centering
\includegraphics[width=8.5cm]{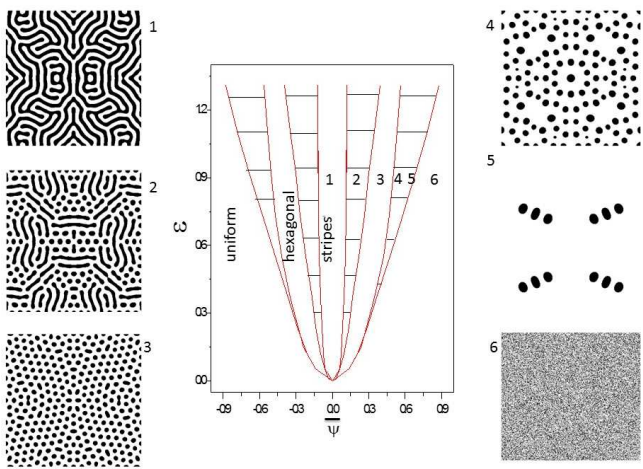}
\caption{A-dimensional phase diagram  function of the nominal composition, $\overline{\psi}$ and the $\epsilon$ parameter. Solid lines represent the limits between different morphology and hatched areas are associated with phase-coexistence domains.
Displayed 2D simulated microstructures (\refeq{eq_Lyapounov},  black for $\eta > 0$, white for $\eta < 0$ ) for domain size, L=200 (reduced units). Values of  $\Delta=0.2$ and $R=3$  correspond to constant $\epsilon=0.86$. From top-left to bottom-right, 1: $\overline{\psi}=0$,2:  $\overline{\psi}= 0.2$, 3: $\overline{\psi}= 0.35$, 4: $\overline{\psi}= 0.5$, 5: $\overline{\psi}= 0.65$, 5: $\overline{\psi}= 0.7$.}
\label{fig_diag-phase-2d} 
\end{figure}

The second step consists in re-writing \refeq{eq_Lyapounov} in reduced units scaling space and time, $l_0=\sqrt{\frac{|a_2|}{a_4 \alpha^2}}$ and $t_0=\frac{1}{\Gamma^{th}(T,\phi)}$, with $\alpha=\frac{\eta_{+} - \eta_{-}}{2}$  and $\eta_{\pm}$, the phase compositions minimizing the homogeneous free-energy density. The Lyapounov functional, $\mathcal{L}$, expression in reduced variables, $r'=\frac{r}{l_0}$, $t'=\frac{t}{t_0}$ and $\eta'(\mathbf{r},t)=\frac{\eta(\mathbf{r},t)}{\alpha}$, is:
\begin{eqnarray}
\label{eq_LSH}
\mathcal{L_{SH}} [\eta']&=& \frac{1}{2} \int \eta'(\mathbf{r'_1,t'}) \left[D(k_0) +\frac{D"(k_0)}{8k_0^2} ({\nabla}^2+{k_0}^2)^2 \right] \nonumber \\ 
& & \eta'(\mathbf{r'_2,t'}) dr'_1 dr'_2 
+  \int \frac{\eta(\mathbf{r'_1,t'})}{4} dr'_1  
\end{eqnarray}
One recognizes in $\mathcal{L_{SH}}$, the Swift-Hohenberg (SH) functional, extensively employed in studies of non equilibrium systems~\cite{Cross93, Elder02, Archer12}, with standard form given by:
\begin{eqnarray}
 \mathcal{L^*_{SH}}[\psi]&  =& \frac{1}{2} \int \psi(\mathbf{r_1}) \left[-\epsilon + (1+\nabla^2)^2 \right] \psi(\mathbf{r_2}) dr_1 dr_2 \nonumber \\
 &+& \int \frac{{\psi(\mathbf{r_1})}^4}{4} d\mathbf{r_1} 
 \label{eq_normal}
 \end{eqnarray}
where $r_i=k_0 r'_i$, $\mathcal{L^*_{SH}}=\mathcal{L_{SH}}'\sqrt{\frac{8}{D''(k_0)k_0^{10}}}$ and $\psi~=~\eta' \sqrt{\frac{8}{D''(k_0)k_0^2}}$. One can write the $\epsilon$ parameter as a function of $\Delta$ and $R$ as: $\epsilon(R,\Delta)=\frac{R^2 \sqrt{\Delta}(2\sqrt{\Delta R^4}-1-R^2)}{(\sqrt{\Delta R^4}-1)^2}$.
In the patterning regime, $\Delta$ values range from $\Delta_{min}$ to  $\Delta_{max}$ given by~\cite{Simeone13}:
\begin{eqnarray}
\label{eq_dmin}
\Delta_{min}=R^{-4}\\
\Delta_{max}=\left({\frac{1+R^2}{2 R^2}} \right)^2
\label{eq_dmax}
\end{eqnarray}
thus implying that, $\epsilon$, is always positive, whereas the minima of $\mathcal{L^*_{SH}}$, defining the phase compositions at the steady state, are functions of the reduced nominal composition of the mixture, $\overline{\psi}$.
\newline
\newline
This is the pivotal result of the present work reducing the irradiation problem into the standard case of the (SH) representation of dynamical systems. As a direct consequence, morphology and composition of the emerging patterns are directly related to the nominal composition $\overline{\psi}$ and the $\epsilon(R,\Delta)$ parameter.
\newline
\newline
For the illustration purpose, a 2D case-study is treated here, whereas the 3D extension of this analysis is straightforward. 
By following the methodology given in \cite{Archer12}, the minima of $\mathcal{L}_{SH}$ are classified for this conservative case in the "phase diagram" displayed in Fig.~\ref{fig_diag-phase-2d}, thus considerably improving the incomplete representation of Fig.~\ref{fig_pattern}. 

\begin{figure}[ht]
\centering 
\includegraphics[width=8cm]{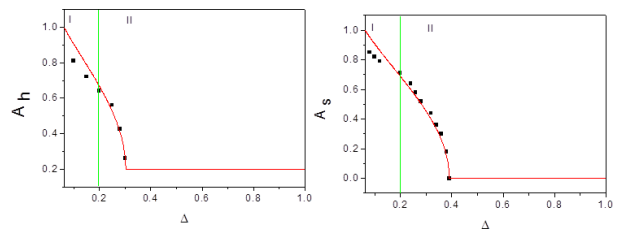}
\caption{Comparison of the solubility limits $A_s$ and $A_h$ (full red line) extracted from minimization of Eq.(\ref{eq_LSH}) and from direct numerical simulations (black squares) for honeycomb (left, $\overline{\psi}=0.25$) and labyrinthine (right, $\overline{\psi}=0$) structures for $R=3$. The dashed line ($\epsilon = 1$), represent  the limitation of the one-mode approximation (I: $\epsilon > 1$, II: $\epsilon < 1$) }
\label{fig_solubility} 
\end{figure}

 In the patterning domain, only three distinct steady-states exist with space distribution of phases that can be identified in the small $\epsilon$ limit (one mode approximation)~\cite{Archer12,Cross93,Elder04}:
\begin{itemize}
    \item a uniform microstructure with $\psi(\mathbf{r})=\overline{\psi}$ (graph 6 in Fig.~\ref{fig_diag-phase-2d}).
    \item a labyrinthine microstructure with composition fluctuations, $\psi(\mathbf{r})=\overline{\psi}+A_s cos(\mathbf{k}  \mathbf{r})$ and wave vectors, $\mathbf{k}=k_0\left( \begin{matrix} 1 \\ 0  \end{matrix}\right)$ and $\mathbf{k}=k_0\left( \begin{matrix} 0 \\ 1  \end{matrix}\right)$ (graph 1 in Fig.~\ref{fig_diag-phase-2d}).
    \item a honeycomb structure with composition fluctuations $\psi(\mathbf{r})=\overline{\psi}+A_h \sum_{j=1}^{3} e^{i \mathbf{k_j} \cdot \mathbf{r}} + c.c. $  and wave vectors, $\mathbf{k_1}=k_0\left( \begin{matrix} 1 \\ 0  \end{matrix}\right)$, $\mathbf{k_2}=k_0\left( \begin{matrix} -1/2 \\ \sqrt{3}/2  \end{matrix}\right)$ and $\mathbf{k_3}=k_0\left( \begin{matrix} -1/2 \\ -\sqrt{3}/2  \end{matrix}\right)$ (graph 3 in Fig.~\ref{fig_diag-phase-2d}).
\end{itemize}
%%%%%%%%%%%%%%%%%%%%%%%%%%%%%%%%%%%%%%%%%%%%%%%%%%%%%%
%%%%%%%%%%%%%%%%%%%%%%%%%%%%%%%%%%%%%%%%%%%%%%%%%%%%%%%%
 As expected, the formation of different microstructures is controlled by the value of $\overline{\psi}$: labyrinthine lamelar stripes at low $|\overline{\psi}|$, a honeycomb structure of spherical precipitates for intermediate  $|\overline{\psi}|$ values and  a homogeneous solid solution at large $|\overline{\psi|}$ values. Moreover, phase-coexistence domains form (hatched areas in fig.~\ref{fig_diag-phase-2d}), combining two different pattern morphologies (graphs 2, 4 and 5 in fig.~\ref{fig_diag-phase-2d}).
It is worth noting that solving numerically Eq.(\ref{eq_Lyapounov}) yields identical results, thus confirming the validity of the approximations made in deriving the $\mathcal{L_{SH}}$ functional. 
\newline
\newline
In parallel to the identification of microstructures, the minimization of $\mathcal{L_{SH}[\psi]}$, provides at the same time the stationary values of the species concentrations specific to each steady-state: $A_s=\frac{2}{3} \sqrt{3 \epsilon-9\overline{\psi}^2}$ and $k_0=1$ for stripes, $A_h=\frac{4}{5}[\overline{\psi}+\frac{\sqrt{15 \epsilon-36 \overline{\psi}^2}}{3}]$ and $k_0= \sqrt{\frac{3}{4}}$ for honeycomb-like patterns. These expressions are explicit functions of the irradiation flux, $\phi$, of the temperature, $T$ and  of the  nominal composition, $\overline{\psi}$, yielding the solubility limits (maximum values of $\psi(\mathbf{r})$). In Fig.~\ref{fig_solubility} these limits are represented as a function of $\Delta$ (full red lines) and reveal remarkably close to the solubility values obtained by solving numerically Eq.~(\ref{eq_Lyapounov}) (full squares).
%%%%%%%%%%%%%%%%%%%%%%%%%%%%%%%%%%%%%%%%%%%%%%%%%%%
%%%%%%%%%%%%%%%%%%%%%%%%%%%%%%%%%%%%%%%%%%%%%%%%%%%%%
Interestingly, the agreement between the two series of data is better in the region labeled II in this figure, where holds the relation, $\epsilon(R,\Delta)<1$. As expected, with increasing $\epsilon$ (decreasing $\Delta$), the the overlap between the corresponding data-sets is not perfect since there, the one-mode approximation fails.
\newline
\newline
As a byproduct of the above results, a PRactical method emerges for studying Irradiated Micro-structures (PRIM), with the power to predict the composition and the symmetry elements of phases without solving explicitly \refeq{eq_Lyapounov}. The method consists in three steps: (i) identifying the set of parameters of the free-energy density, ${a_2}$, ${a_3}$, ${a_4}$, such as to reproduce  the experimental phase diagram and the interfacial stiffness coefficient, $\kappa$, from experiments or via MD and Monte-Carlo (MC) simulations~\cite{Demange15}, (ii) estimating the kinetic enhancement factor, $\Delta$, from knowledge of the thermal mobility and of the relocation distance of species $R$ and (iii) locating in the a-dimensional phase diagram of Fig.~\ref{fig_diag-phase-2d} the steady-state patterns and computing the phase compositions, from the reduced values of $\epsilon$ and $\overline{\psi}$, obtained in the preceding two steps.

\begin{figure}[ht]
\centering
\includegraphics[width=8cm]{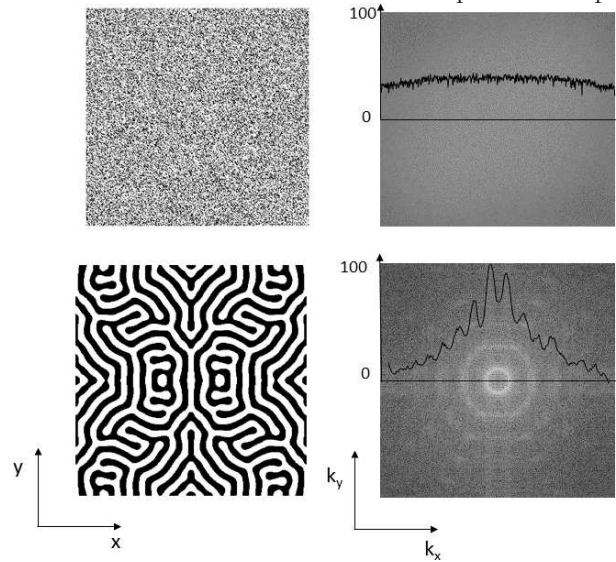}
\caption{Random $Ag_{0.39}Cu_{0.61}$ mixture ($\overline{\psi}=0$), before (top) and after (bottom) irradiation with 1 MeV Kr ions at flux $\phi=6 \times 10^{12} cm^{-2}s^{-1}$ ($R=3$ and $\Delta=0.2$). Initial and steady state 2D distributions of species (left-hand side) and the corresponding structure factor 2D representations (right-hand side). It should be noted that in the steady state the structure factor is radially symmetric and sharply peaked around $k_0$ (white ring, bottom right image). The full line in these representations is the profile of the structure factor along the $k_x$ component of the wave vector at  $k_y=0$.}
\label{fig_calcul} 
\end{figure}

For the illustration purpose, the PRIM method is here used to predict the long-time evolution of an homogeneous Ag-Cu mixture ($\overline{c}_{Ag}=0.39$) irradiated by 1 MeV Kr ions at flux value, $\phi=6 \times 10^{12} cm^{-2}s^{-1}$ at $T=440 K$. The free-energy density parameters, ${a_2}$, ${a_3}$, ${a_4}$, have been determined as indicated above in (i) and are further detailed in~\cite{Subram93},  whereas the interfacial stiffness coefficient, $\kappa$, has been estimated by fitting on to the grand-canonical MC prediction of the species composition profiles across Ag/Cu (100) semi-coherent interfaces the predictions via \refeq{eq_CH} including the Ginzburg term~\cite{Demange15}. The kinetic enhancement factor has been extracted from MD simulations and experimental thermal mobility values~\cite{Enrique04,Simeone10,Subram93}. Therefrom, the values are obtained, $R=3$, $\Delta=0.2$, leading to $\epsilon=0.86$ and $\overline{\psi}=0$, which correspond to the 2D steady-state pattern shown in the left-hand side of Fig.~\ref{fig_calcul} according to the a-dimensional phase diagram in Fig.~\ref{fig_diag-phase-2d}. 
Worth noting, the associated 2D structure factor is sharply peaked as expected (right-hand side, bottom of Fig.~\ref{fig_calcul}). The same steady-state microstructure is also predicted  for the equimolar AgCu mixture studied by Enrique et al.~\cite{Enrique04} for which, $\overline{\psi} \neq 0$ (Fig.~\ref{fig_calcul}).

Modelling the effects of binary mixtures submitted to irradiation and predicting the steady-state patterns thereby produced has been the subject of several contributions in recent years. Among these, Martin~\cite{Martin84} has proposed a theoretical analysis that led to the concept of 'effective temperature' according which irradiation acts simply as an excess temperature enhancing the evolution of the mixture and producing dynamical steady-states. However, this analysis has revealed unable predicting the ordered steady-state patterns the present work shows triggered by the irradiation~\cite{Enrique00,Simeone13}. Subsequent contributions to the subject~\cite{Enrique04} have not identified the crucial role of the nominal composition of the mixture in determining the multiplicity of steady-state irradiation patterns and the corresponding compositions of phases that the present work has firmly established.
%%%%%%%%%%%%%%%%%%%%%%%%%%%%%%%%%%%%%%%%%%%%%%%%%%%%%%%%%%%%%%%
%%%%%%%%%%%%%%%%%%%%%%%%%%%%%%%%%%%%%%%%%%%%%%%%%%%%%%%%%%%%%%%%%

In summary, the present work shows that irradiation phase diagrams should be drawn in the three dimensional space spanned by the nominal composition of the considered mixture, the relocation average distance of species and the kinetic enhancement factor combining thermal and irradiation triggered mobilities. The theoretical developments presented here allow for predicting irradiation patterns and phases compositions via an a-dimensional phase diagram conveniently integrated within an operational method (PRIM), bypassing the need of solving the evolution equations for the case of interest. Applying the PRIM method in experimental studies, would facilitate identifying the characteristic features of irradiation microstructures, which might constitute a decisive contribution in this area crucially lacking of experimental support.

Ongoing and future work focus respectively on experimental investigations of mixtures decomposing in presence of irradiation with ions and on investigation of the relative stability of irradiation-triggered steady states (noise effects)~\cite{Beauford15}.

We thank A. Forestier and N. Ofori-Opoku for helpful remarks.

%\bibliography{biblio.bib}

\begin{thebibliography}{21}%
\makeatletter
\providecommand \@ifxundefined [1]{%
 \@ifx{#1\undefined}
}%
\providecommand \@ifnum [1]{%
 \ifnum #1\expandafter \@firstoftwo
 \else \expandafter \@secondoftwo
 \fi
}%
\providecommand \@ifx [1]{%
 \ifx #1\expandafter \@firstoftwo
 \else \expandafter \@secondoftwo
 \fi
}%
\providecommand \natexlab [1]{#1}%
\providecommand \enquote  [1]{``#1''}%
\providecommand \bibnamefont  [1]{#1}%
\providecommand \bibfnamefont [1]{#1}%
\providecommand \citenamefont [1]{#1}%
\providecommand \href@noop [0]{\@secondoftwo}%
\providecommand \href [0]{\begingroup \@sanitize@url \@href}%
\providecommand \@href[1]{\@@startlink{#1}\@@href}%
\providecommand \@@href[1]{\endgroup#1\@@endlink}%
\providecommand \@sanitize@url [0]{\catcode `\\12\catcode `\$12\catcode
  `\&12\catcode `\#12\catcode `\^12\catcode `\_12\catcode `\%12\relax}%
\providecommand \@@startlink[1]{}%
\providecommand \@@endlink[0]{}%
\providecommand \url  [0]{\begingroup\@sanitize@url \@url }%
\providecommand \@url [1]{\endgroup\@href {#1}{\urlprefix }}%
\providecommand \urlprefix  [0]{URL }%
\providecommand \Eprint [0]{\href }%
\providecommand \doibase [0]{http://dx.doi.org/}%
\providecommand \selectlanguage [0]{\@gobble}%
\providecommand \bibinfo  [0]{\@secondoftwo}%
\providecommand \bibfield  [0]{\@secondoftwo}%
\providecommand \translation [1]{[#1]}%
\providecommand \BibitemOpen [0]{}%
\providecommand \bibitemStop [0]{}%
\providecommand \bibitemNoStop [0]{.\EOS\space}%
\providecommand \EOS [0]{\spacefactor3000\relax}%
\providecommand \BibitemShut  [1]{\csname bibitem#1\endcsname}%
\let\auto@bib@innerbib\@empty
%</preamble>
\bibitem [{\citenamefont {Adda}\ \emph {et~al.}(1975)\citenamefont {Adda},
  \citenamefont {Beyeler},\ and\ \citenamefont {Brebec}}]{Adda75}%
  \BibitemOpen
  \bibfield  {author} {\bibinfo {author} {\bibfnamefont {Y.}~\bibnamefont
  {Adda}}, \bibinfo {author} {\bibfnamefont {M.}~\bibnamefont {Beyeler}}, \
  and\ \bibinfo {author} {\bibfnamefont {G.}~\bibnamefont {Brebec}},\
  }\href@noop {} {\bibfield  {journal} {\bibinfo  {journal} {Thin Solid Films}\
  }\textbf {\bibinfo {volume} {25}},\ \bibinfo {pages} {S28} (\bibinfo {year}
  {1975})}\BibitemShut {NoStop}%
\bibitem [{\citenamefont {Barbu}\ \emph {et~al.}(1980)\citenamefont {Barbu},
  \citenamefont {Martin},\ and\ \citenamefont {Chamberod}}]{Barbu80}%
  \BibitemOpen
  \bibfield  {author} {\bibinfo {author} {\bibfnamefont {A.}~\bibnamefont
  {Barbu}}, \bibinfo {author} {\bibfnamefont {G.}~\bibnamefont {Martin}}, \
  and\ \bibinfo {author} {\bibfnamefont {A.}~\bibnamefont {Chamberod}},\
  }\href@noop {} {\bibfield  {journal} {\bibinfo  {journal} {J. Appl. Phys.}\
  }\textbf {\bibinfo {volume} {51}},\ \bibinfo {pages} {126192} (\bibinfo
  {year} {1980})}\BibitemShut {NoStop}%
\bibitem [{\citenamefont {Enrique}\ and\ \citenamefont
  {Bellon}(2000)}]{Enrique00}%
  \BibitemOpen
  \bibfield  {author} {\bibinfo {author} {\bibfnamefont {R.~A.}\ \bibnamefont
  {Enrique}}\ and\ \bibinfo {author} {\bibfnamefont {P.}~\bibnamefont
  {Bellon}},\ }\href@noop {} {\bibfield  {journal} {\bibinfo  {journal} {Phys.
  Rev. Lett.}\ }\textbf {\bibinfo {volume} {84}},\ \bibinfo {pages} {2885}
  (\bibinfo {year} {2000})}\BibitemShut {NoStop}%
\bibitem [{\citenamefont {Simeone}\ \emph {et~al.}(2013)\citenamefont
  {Simeone}, \citenamefont {Demange},\ and\ \citenamefont
  {Luneville}}]{Simeone13}%
  \BibitemOpen
  \bibfield  {author} {\bibinfo {author} {\bibfnamefont {D.}~\bibnamefont
  {Simeone}}, \bibinfo {author} {\bibfnamefont {G.}~\bibnamefont {Demange}}, \
  and\ \bibinfo {author} {\bibfnamefont {L.}~\bibnamefont {Luneville}},\
  }\href@noop {} {\bibfield  {journal} {\bibinfo  {journal} {Phys. Rev. E}\
  }\textbf {\bibinfo {volume} {88}},\ \bibinfo {pages} {032116} (\bibinfo
  {year} {2013})}\BibitemShut {NoStop}%
\bibitem [{\citenamefont {Demange}\ \emph {et~al.}(2017)\citenamefont
  {Demange}, \citenamefont {Luneville}, \citenamefont {Pontikis},\ and\
  \citenamefont {Simeone}}]{Demange15}%
  \BibitemOpen
  \bibfield  {author} {\bibinfo {author} {\bibfnamefont {G.}~\bibnamefont
  {Demange}}, \bibinfo {author} {\bibfnamefont {L.}~\bibnamefont {Luneville}},
  \bibinfo {author} {\bibfnamefont {V.}~\bibnamefont {Pontikis}}, \ and\
  \bibinfo {author} {\bibfnamefont {D.}~\bibnamefont {Simeone}},\ }\href@noop
  {} {\bibfield  {journal} {\bibinfo  {journal} {J. Appl. Phys.}\ }\textbf
  {\bibinfo {volume} {121}},\ \bibinfo {pages} {125108} (\bibinfo {year}
  {2017})}\BibitemShut {NoStop}%
\bibitem [{\citenamefont {Subramanian}\ and\ \citenamefont
  {Perepezko}(1993)}]{Subram93}%
  \BibitemOpen
  \bibfield  {author} {\bibinfo {author} {\bibfnamefont {P.}~\bibnamefont
  {Subramanian}}\ and\ \bibinfo {author} {\bibfnamefont {J.}~\bibnamefont
  {Perepezko}},\ }\href@noop {} {\bibfield  {journal} {\bibinfo  {journal}
  {Journal of Phase Equilibrium}\ }\textbf {\bibinfo {volume} {14}},\ \bibinfo
  {pages} {62} (\bibinfo {year} {1993})}\BibitemShut {NoStop}%
\bibitem [{\citenamefont {Sigmund}\ and\ \citenamefont
  {Gras-Marti}(1981)}]{Sigmund81}%
  \BibitemOpen
  \bibfield  {author} {\bibinfo {author} {\bibfnamefont {P.}~\bibnamefont
  {Sigmund}}\ and\ \bibinfo {author} {\bibfnamefont {A.}~\bibnamefont
  {Gras-Marti}},\ }\href@noop {} {\bibfield  {journal} {\bibinfo  {journal}
  {Nucl. Inst. and Methods B}\ }\textbf {\bibinfo {volume} {182}},\ \bibinfo
  {pages} {211 } (\bibinfo {year} {1981})}\BibitemShut {NoStop}%
\bibitem [{\citenamefont {Khatchaturyan}(1983)}]{Khatchaturyan83}%
  \BibitemOpen
  \bibfield  {author} {\bibinfo {author} {\bibfnamefont {A.~G.}\ \bibnamefont
  {Khatchaturyan}},\ }\href@noop {} {\emph {\bibinfo {title} {Theory of
  structural transformation in solids}}}\ (\bibinfo  {publisher} {Wiley
  Interscience},\ \bibinfo {year} {1983})\BibitemShut {NoStop}%
\bibitem [{\citenamefont {Tol{\'e}dano}\ and\ \citenamefont
  {Dmitriev}(1996)}]{Toledano96}%
  \BibitemOpen
  \bibfield  {author} {\bibinfo {author} {\bibfnamefont {P.}~\bibnamefont
  {Tol{\'e}dano}}\ and\ \bibinfo {author} {\bibfnamefont {V.}~\bibnamefont
  {Dmitriev}},\ }\href@noop {} {\emph {\bibinfo {title} {Reconstructive phase
  transitions: in crystals and quasicrystals}}}\ (\bibinfo  {publisher} {World
  Scientific},\ \bibinfo {year} {1996})\BibitemShut {NoStop}%
\bibitem [{\citenamefont {Gras-Marti}\ and\ \citenamefont
  {Sigmund}(1981)}]{GrasMarti81}%
  \BibitemOpen
  \bibfield  {author} {\bibinfo {author} {\bibfnamefont {A.}~\bibnamefont
  {Gras-Marti}}\ and\ \bibinfo {author} {\bibfnamefont {P.}~\bibnamefont
  {Sigmund}},\ }\href@noop {} {\bibfield  {journal} {\bibinfo  {journal} {Nucl.
  Inst. and Methods B}\ }\textbf {\bibinfo {volume} {180}},\ \bibinfo {pages}
  {211 } (\bibinfo {year} {1981})}\BibitemShut {NoStop}%
\bibitem [{\citenamefont {Simeone}\ and\ \citenamefont
  {Luneville}(2010)}]{Simeone10}%
  \BibitemOpen
  \bibfield  {author} {\bibinfo {author} {\bibfnamefont {D.}~\bibnamefont
  {Simeone}}\ and\ \bibinfo {author} {\bibfnamefont {L.}~\bibnamefont
  {Luneville}},\ }\href@noop {} {\bibfield  {journal} {\bibinfo  {journal}
  {Phys. Rev. E}\ }\textbf {\bibinfo {volume} {81}},\ \bibinfo {pages} {021115}
  (\bibinfo {year} {2010})}\BibitemShut {NoStop}%
\bibitem [{\citenamefont {Luneville}\ \emph {et~al.}(2016)\citenamefont
  {Luneville}, \citenamefont {Mallick}, \citenamefont {Pontikis},\ and\
  \citenamefont {Simeone}}]{Luneville16}%
  \BibitemOpen
  \bibfield  {author} {\bibinfo {author} {\bibfnamefont {L.}~\bibnamefont
  {Luneville}}, \bibinfo {author} {\bibfnamefont {K.}~\bibnamefont {Mallick}},
  \bibinfo {author} {\bibfnamefont {V.}~\bibnamefont {Pontikis}}, \ and\
  \bibinfo {author} {\bibfnamefont {D.}~\bibnamefont {Simeone}},\ }\href@noop
  {} {\bibfield  {journal} {\bibinfo  {journal} {Phys. Rev. E}\ }\textbf
  {\bibinfo {volume} {94}},\ \bibinfo {pages} {052126} (\bibinfo {year}
  {2016})}\BibitemShut {NoStop}%
\bibitem [{\citenamefont {Glotzer}\ \emph {et~al.}(1995)\citenamefont
  {Glotzer}, \citenamefont {Di~Marzio},\ and\ \citenamefont
  {Muthukumar}}]{Glotzer95}%
  \BibitemOpen
  \bibfield  {author} {\bibinfo {author} {\bibfnamefont {S.}~\bibnamefont
  {Glotzer}}, \bibinfo {author} {\bibfnamefont {E.}~\bibnamefont {Di~Marzio}},
  \ and\ \bibinfo {author} {\bibfnamefont {M.}~\bibnamefont {Muthukumar}},\
  }\href@noop {} {\bibfield  {journal} {\bibinfo  {journal} {Phys. Rev. Lett.}\
  }\textbf {\bibinfo {volume} {74}},\ \bibinfo {pages} {2034} (\bibinfo {year}
  {1995})}\BibitemShut {NoStop}%
\bibitem [{\citenamefont {Jin}\ and\ \citenamefont
  {Katchaturyan}(2006)}]{Jin06}%
  \BibitemOpen
  \bibfield  {author} {\bibinfo {author} {\bibfnamefont {Y.}~\bibnamefont
  {Jin}}\ and\ \bibinfo {author} {\bibfnamefont {A.}~\bibnamefont
  {Katchaturyan}},\ }\href@noop {} {\bibfield  {journal} {\bibinfo  {journal}
  {Journal of Applied Physics}\ }\textbf {\bibinfo {volume} {100}},\ \bibinfo
  {pages} {013519} (\bibinfo {year} {2006})}\BibitemShut {NoStop}%
\bibitem [{\citenamefont {Cross}\ and\ \citenamefont
  {Hohenberg}(1993)}]{Cross93}%
  \BibitemOpen
  \bibfield  {author} {\bibinfo {author} {\bibfnamefont {M.~C.}\ \bibnamefont
  {Cross}}\ and\ \bibinfo {author} {\bibfnamefont {P.~C.}\ \bibnamefont
  {Hohenberg}},\ }\href@noop {} {\bibfield  {journal} {\bibinfo  {journal}
  {Rev. Mod. Phys.}\ }\textbf {\bibinfo {volume} {65}},\ \bibinfo {pages} {851}
  (\bibinfo {year} {1993})}\BibitemShut {NoStop}%
\bibitem [{\citenamefont {Elder}\ \emph {et~al.}(2002)\citenamefont {Elder},
  \citenamefont {Katakowski}, \citenamefont {Haataja},\ and\ \citenamefont
  {Grant}}]{Elder02}%
  \BibitemOpen
  \bibfield  {author} {\bibinfo {author} {\bibfnamefont {K.}~\bibnamefont
  {Elder}}, \bibinfo {author} {\bibfnamefont {M.}~\bibnamefont {Katakowski}},
  \bibinfo {author} {\bibfnamefont {M.}~\bibnamefont {Haataja}}, \ and\
  \bibinfo {author} {\bibfnamefont {M.}~\bibnamefont {Grant}},\ }\href@noop {}
  {\bibfield  {journal} {\bibinfo  {journal} {Phys. Rev. Lett.}\ }\textbf
  {\bibinfo {volume} {88}},\ \bibinfo {pages} {245701} (\bibinfo {year}
  {2002})}\BibitemShut {NoStop}%
\bibitem [{\citenamefont {Archer}\ \emph {et~al.}(2012)\citenamefont {Archer},
  \citenamefont {Robbins}, \citenamefont {Thiele},\ and\ \citenamefont
  {Knobloch}}]{Archer12}%
  \BibitemOpen
  \bibfield  {author} {\bibinfo {author} {\bibfnamefont {A.}~\bibnamefont
  {Archer}}, \bibinfo {author} {\bibfnamefont {M.}~\bibnamefont {Robbins}},
  \bibinfo {author} {\bibfnamefont {U.}~\bibnamefont {Thiele}}, \ and\ \bibinfo
  {author} {\bibfnamefont {E.}~\bibnamefont {Knobloch}},\ }\href@noop {}
  {\bibfield  {journal} {\bibinfo  {journal} {Phys Rev E}\ }\textbf {\bibinfo
  {volume} {46}},\ \bibinfo {pages} {31603} (\bibinfo {year}
  {2012})}\BibitemShut {NoStop}%
\bibitem [{\citenamefont {Elder}\ and\ \citenamefont {Grant}(2004)}]{Elder04}%
  \BibitemOpen
  \bibfield  {author} {\bibinfo {author} {\bibfnamefont {K.}~\bibnamefont
  {Elder}}\ and\ \bibinfo {author} {\bibfnamefont {M.}~\bibnamefont {Grant}},\
  }\href@noop {} {\bibfield  {journal} {\bibinfo  {journal} {Phys. Rev. E}\
  }\textbf {\bibinfo {volume} {70}},\ \bibinfo {pages} {051605} (\bibinfo
  {year} {2004})}\BibitemShut {NoStop}%
\bibitem [{\citenamefont {Enrique}\ and\ \citenamefont
  {Bellon}(2004)}]{Enrique04}%
  \BibitemOpen
  \bibfield  {author} {\bibinfo {author} {\bibfnamefont {R.}~\bibnamefont
  {Enrique}}\ and\ \bibinfo {author} {\bibfnamefont {P.}~\bibnamefont
  {Bellon}},\ }\href@noop {} {\bibfield  {journal} {\bibinfo  {journal} {Phys.
  Rev. B}\ }\textbf {\bibinfo {volume} {70}},\ \bibinfo {pages} {224106}
  (\bibinfo {year} {2004})}\BibitemShut {NoStop}%
\bibitem [{\citenamefont {Martin}(1984)}]{Martin84}%
  \BibitemOpen
  \bibfield  {author} {\bibinfo {author} {\bibfnamefont {G.}~\bibnamefont
  {Martin}},\ }\href@noop {} {\bibfield  {journal} {\bibinfo  {journal} {Phys.
  Rev. B}\ }\textbf {\bibinfo {volume} {30}},\ \bibinfo {pages} {53} (\bibinfo
  {year} {1984})}\BibitemShut {NoStop}%
\bibitem [{\citenamefont {Beauford}\ \emph {et~al.}(2015)\citenamefont
  {Beauford}, \citenamefont {Vallet}, \citenamefont {Nicolai},\ and\
  \citenamefont {Bardot}}]{Beauford15}%
  \BibitemOpen
  \bibfield  {author} {\bibinfo {author} {\bibfnamefont {M.}~\bibnamefont
  {Beauford}}, \bibinfo {author} {\bibfnamefont {M.}~\bibnamefont {Vallet}},
  \bibinfo {author} {\bibfnamefont {J.}~\bibnamefont {Nicolai}}, \ and\
  \bibinfo {author} {\bibfnamefont {J.}~\bibnamefont {Bardot}},\ }\href@noop {}
  {\bibfield  {journal} {\bibinfo  {journal} {Journal of Applied Physics}\
  }\textbf {\bibinfo {volume} {118}},\ \bibinfo {pages} {205904} (\bibinfo
  {year} {2015})}\BibitemShut {NoStop}%
\end{thebibliography}
\end{document}